\documentstyle[12pt,bezier]{article}
\begin{document}
\def \inbar{\vrule height1.5ex width.4pt depth0pt}
\def \xC{\relax\hbox{\kern.25em$\inbar\kern-.3em{\rm C}$}}
\def \xR{\relax{\rm I\kern-.18em R}}
\newcommand{\xZ}{Z \hspace{-.08in}Z}
\newcommand{\xbe}{\begin{equation}}
\newcommand{\xee}{\end{equation}}
\newcommand{\xbea}{\begin{eqnarray}}
\newcommand{\xeea}{\end{eqnarray}}
\newcommand{\xnn}{\nonumber}
\newcommand{\xkt}{\rangle}
\newcommand{\xbr}{\langle}
\newcommand{\xlll}{\left( }
\newcommand{\xrrr}{\right)}
\newcommand{\xcun}{\mbox{\footnotesize${\cal N}$}}
\newcommand{\cun}{{\mbox{\footnotesize${\cal N}$}}}
\title{Non-Abelian Geometric Phase, Floquet Theory, and 
Periodic Dynamical Invariants}
\author{Ali Mostafazadeh\thanks{E-mail address: 
amostafazadeh@ku.edu.tr}\\ \\
Department of Mathematics, Ko\c{c} University,\\
Istinye 80860, Istanbul, TURKEY}
\date{ }
\maketitle

\begin{abstract}
For a periodic Hamiltonian, periodic dynamical invariants may be used
to obtain non-degenerate cyclic states. This observation is generalized
to the degenerate cyclic states, and the relation between the periodic
dynamical invariants and the Floquet decompositions of the time-evolution 
operator is elucidated. In particular, a necessary condition for
the occurrence of cyclic non-adiabatic non-Abelian geometrical phase is derived.
Degenerate cyclic states are obtained for a magnetic dipole interacting with 
a precessing magnetic field. 

\end{abstract}
\vspace{3mm}
\vspace{3mm}

\baselineskip=24pt

\section{Introduction}

Since Berry's article \cite{berry1984} on adiabatic geometric phase, there
has been a growing number of publications on the subject. The most notable
developments have been the characterization of Berry's adiabatic phase as
the holonomy of a spectral bundle \cite{simon}, the discovery of the 
non-Abelian \cite{wi-ze} and classical \cite{hannay} analogues of Berry's 
adiabatic phase, and its generalization to non-adiabatic 
cyclic \cite{aa} and even non-cyclic evolutions \cite{noncyclic}. The main
reason for the enormous excitement generated by Berry's finding
\cite{berry1984}
has been the wide range of its application in different areas of physics
\cite{applications} and its rich mathematical structure \cite{math}. 

Indeed, it is quite surprising that geometric phases were not discovered 
much earlier. There are a few older papers \cite{older} in the literature where
the authors come very close to discovering the geometric phase. Perhaps one 
of the most important of these is the classic 1969 paper of Lewis and Riesenfeld 
\cite{le-ri} on dynamical invariants. The correspondence of Berry's phase and 
Lewis's phase has been pointed out by Morales \cite{morales}. More recently, 
Monteoliva, Korsch and N\'u$\tilde{\rm n}$es \cite{mkn} showed that if a 
periodic invariant operator $I(t)$ with non-degenerate spectrum was known,
then one could obtain all the pure cyclic states as the eigenstates of $I(0)$
and compute the corresponding geometric phases directly in terms of $I(t)$. 
These authors also pointed out that for a periodic Hamiltonian $H(t)$ their 
method was more convenient than performing a Floquet decomposition 
$Z(t)e^{iMt}$ of the time-evolution operator $U(t)$ and obtaining the pure 
cyclic states as the eigenstates of the operator $M$, as originally suggested 
by Moore and Stedman \cite{m-s,moore} (See also the paper of Furman \cite{fur}.) 

In the present paper, we shall generalize the results of Monteoliva, et al
\cite{mkn} to degenerate cyclic evolutions. This generalization involves
the analysis of the relationship between the periodic dynamical invariants
$I(t)$ and the Floquet operators $Z(t)$ and $M$. In particular, we obtain
a necessary condition for the occurrence of non-adiabatic non-Abelian 
geometrical phases factors, and show that a non-degenerate Hamiltonian may support
degenerate cyclic evolutions. A simple example is provided by the quantum
dynamics of a magnetic dipole interacting with a precessing magnetic field.

The organization of the paper is as follows. In section~2, we recall
the basic results of the Floquet theory for periodic Hamiltonians. In 
section~3, we discuss the relevance of the Lewis-Riesenfeld theory of 
dynamical invariants to the geometric phase. In particular, we derive
the expression for the non-Abelian Lewis's phase and demonstrate its
coincidence with the non-adiabatic non-Abelian geometric phase
\cite{anandan}. 
In section~4, we offer a generalization of the results of Monteoliva, et al
\cite{mkn} to degenerate cyclic evolutions and discuss a characterization
of degenerate cyclic states. In section~5, we apply these results to obtain a 
degenerate cyclic state of a magnetic dipole (of spin $j=1$) interacting 
with a precessing magnetic field. In section~6, we present a summary of 
our results and conclude the paper with some final remarks.

In the following we shall set $\hbar=1$, reperesent quantum states
with projection operators, and assume that
	\begin{itemize}
	\item[1)] the Hamiltonian $H(t)$ is a $T$-periodic Hermitian operator 
	with a discrete spectrum, and
	\item[2)] during the evolution of the system there is no level-crossings.
	In particular, the degree of degeneracy of the eigenvalues of the
	relevant operators will not depend on time.
	\end{itemize}

\section{Floquet Theory}

A classic result of the Floquet theory \cite{floquet} is that for a periodic
Hamiltonian $H(t)$ with period $T$, the time-evolution operator $U(t)$
can be expressed as
	\xbe	
	U(t)=Z(t)e^{iMt}\;,
	\label{flo}
	\xee
where $Z(t)$ is a unitary $T$-periodic operator, i.e. $Z(T)=Z(0)=1$,
and $M$ is a Hermitian operator. Clearly, $U(T)=e^{iMT}$ and the cyclic 
states of the system with period $T$ are the eigenstates of 
$M$.\footnote{In some cases, there are special values of $T$ for which
an eigenvector of $e^{iMT}$ is not an eigenvector of $M$. For example
let $T=4\pi$ and $M=J_3$, where $J_3$ is the $z$-component of the 
angular momentum operator. Then $e^{iMT}=e^{4\pi iJ_3}$ is the identity
operator, and any vector $|\psi\xkt$ is an eigenvector of $e^{iMT}$.
Taking $|\psi\xkt$ to be the sum of two eigenvector of $J_3$, we have
an example of a vector which is an eigenvector of $e^{iMT}$ but not
an eigenvector of $M$. In this article we shall not consider these
special cases.}

We shall assume that
	\begin{itemize}
	\item[1)] $M$ has a discrete spectrum,
	\item[2)] the eigenvalues $\mu_n$ of $M$ are $m_n$-fold degenerate,
	and the corresponding eigenvectors $|\mu_n,a\xkt$ with $a\in\{1,2,\cdots,
	m_n\}$ form a complete orthonormal set of basis vectors of the Hilbert space
	${\cal H}$.
	\end{itemize}

In view of Eq.~(\ref{flo}), if $ |\psi(t)\xkt $ is the solution of the 
Schr\"odinger equation,
	\xbe
	i\frac{d}{dt} |\psi(t)\xkt=H(t) |\psi(t)\xkt\;,
	\label{sch-eq}
	\xee
with the initial condition:
	\xbe
	|\psi(0)\xkt=|\mu_n,a\xkt\;,
	\label{ini-condi}
	\xee
then	
	\xbe
	|\psi(T)\xkt=U(T) |\mu_n,a\xkt =e^{i\mu_nT} |\mu_n,a\xkt\;.
	\label{cyc}
	\xee
Therefore the total phase angle associated with the cyclic state vector $|\mu_n,a\xkt$
is given by
	\xbe
	\alpha_n=\mu_nT\;.
	\label{total-ph}
	\xee
This phase angle consists of a dynamical part $\delta_n$ and a geometric part 
$\gamma_n$ (i.e., $\alpha_n=\delta_n+\gamma_n$) which are expressed in 
the form \cite{aa,m-s,pra2}:
	\xbea
	\delta_n&=&-\int_0^T\xbr\psi(t)|H(t)|\psi(t)\xkt dt\;,
	\label{delta}\\
	\gamma_n &=&i\int_0^T\xbr\phi(t)|\frac{d}{dt}|\phi(t)\xkt dt\;,
	\label{gamma}
	\xeea
where $|\phi(t)\xkt$ is a single-valued state vector defining the same pure
state as $|\psi(t)\xkt$, \cite{aa,bohm-qm}. A particular choice for 
$|\phi(t)\xkt$ is $Z(t)|\mu_n,a\xkt$, \cite{m-s}. Hence,
	\xbe
	\gamma_n=i\int_0^T\xbr\mu_n,a|Z^\dagger(t)\frac{d}{dt}Z(t)
	|\mu_n,a\xkt dt\;.
	\label{gamma-z}
	\xee

Note that in general $m_n>1$, i.e., $\mu_n$ is degenerate. Nevertheless, the 
associated eigenvectors $|\mu_n,a\xkt$ undergo cyclic evolutions, and the 
corresponding geometric phase factors are Abelian. 

\section{Periodic Dynamical Invariants and Degenerate Cyclic Evolutions}

By definition a dynamical invariant $I(t)$ is a solution of
	\xbe
	\frac{dI(t)}{dt}=i\left[I(t),H(t)\right]\;.
	\label{I-eq}
	\xee
In the following we shall assume that
	\begin{itemize}
	\item[1)] $I(t)$ is a Hermitian operator with a discrete spectrum,
	\item[2)] the eigenvalues $\lambda_n$ of $I(t)$ are $l_n$-fold degenerate,
	and the corresponding eigenvectors $|\lambda_n,a;t\xkt$ with $a\in\{1,2,\cdots,
	l_n\}$ form a complete orthonormal set of basis vectors of the Hilbert space
	${\cal H}$.
	\end{itemize}
Note that the eigenvalue equation
	\xbe
	I(t)|\lambda_n,a;t\xkt=\lambda_n|\lambda_n,a;t\xkt\;,
	\label{eg-va-eq-I}
	\xee
determines the eigenvectors $|\lambda_n,a;t\xkt$ uniquely upto possibly time-dependent
unitary transformations $u$ which act on the degeneracy subspace 
	\xbe
	{\cal H}_{\lambda_n}(t):={\rm Span}\left\{|\lambda_n,1;t\xkt,|\lambda_n,2;t\xkt,
	\cdots,|\lambda_n,l_n;t\xkt\right\}
	\label{H_lambda}
	\xee
associated with the eigenvalue $\lambda_n$. In other words, one may choose another 
set of complete orthonormal eigenvectors $|\lambda_n,a;t\xkt'$ of $I(t)$ which are 
related to $|\lambda_n,a;t\xkt$ according to
	\xbe
	|\lambda_n,b;t\xkt'=\sum_{a=1}^{l_n}|\lambda_n,a;t\xkt u_{ab}(t) \;,
	\label{u-trans}
	\xee
where $u_{ab}(t)$ are entries of an $l_n\times l_n$ unitary matrix $u$. 

Now following Lewis and Riesenfeld \cite{le-ri}, let us differentiate both sides 
of Eq.~(\ref{eg-va-eq-I}), take the inner product of both sides of the resulting 
equation with $|\lambda_m,b;t\xkt$, for some $m$ and $b\in\{1,\cdots,l_m\}$, and use 
Eqs.~(\ref{I-eq}) and (\ref{eg-va-eq-I}) to simplify the result. This leads to 
	\xbe
	(\lambda_n-\lambda_m)\left[\xbr\lambda_m,b;t|H|\lambda_n,a;t\xkt-i
	\xbr\lambda_m,b;t|\frac{d}{dt}|\lambda_n,a;t\xkt\right]=
	i\delta_{mn}\delta_{ab}\frac{d}{dt}\lambda_n\;,
	\label{lewis}
	\xee
where $\delta$'s are the Kronecker delta functions. Eq.~(\ref{lewis})
implies:
	\begin{itemize}
	\item[1)] Eigenvalues $\lambda_n$ do not depend on time.
	\item[2)] Eigenvectors $|\lambda_n,a;t\xkt$ satisfy:
	\xbe
	\xbr\lambda_m,b;t|H|\lambda_n,a;t\xkt-i\xbr\lambda_m,b;t|\frac{d}{dt}
	|\lambda_n,a;t\xkt=0,~~~{\rm for}~~~m\neq n\;.
	\label{condi-1}
	\xee
	\end{itemize}
Clearly if Eq.~(\ref{condi-1}) is also satisfied for $m=n$, then
	\xbe
	\xbr\lambda_m,b;t|(H-i\frac{d}{dt})|\lambda_n,a;t\xkt=0,
	~~~{\rm for~all}~~m~~{\rm and}~~b\;,
	\label{condi-2}
	\xee
and $|\lambda_n,a;t\xkt$ is a solution of the Schr\"odinger equation (\ref{sch-eq}).

A central result of Lewis and Riesenfeld \cite{le-ri} is that although 
Eq.~(\ref{condi-2}) may not be satisfied, there are unitary transformations of
the form (\ref{u-trans}) which map $|\lambda_n,a;t\xkt$ to a new set of eigenvectors
$|\lambda_n,a;t\xkt'$ of $I(t)$ which do satisfy Eq.~(\ref{condi-2}) and provide
solutions of the Schr\"odinger equation. Lewis and Riesenfeld do not in fact derive
the defining equation for the matrix $u$. They suffice to say that this matrix
may be diagonalized and obtain the equation satisfied by its eigenvalues, the Lewis
phases.

In order to obtain the appropriate unitary transformation $u$ for a given eigenvalue
$\lambda_n$, we demand that Eq.~(\ref{condi-2}) be fulfilled for the primed 
eigenvectors. Setting $m=n$, using Eqs.~(\ref{u-trans}) and~(\ref{condi-1}), and
simplifying the resulting expression, we obtain
	\xbe
	i\frac{d}{dt}u(t)=\Delta(t)u(t)\;,
	\label{u=}
	\xee
where $\Delta(t)$ is an $l_n\times l_n$ matrix with entries 
	\xbea
	\Delta_{ab}(t)&:=&{\cal E}_{ab}(t)-{\cal A}_{ab}(t)\;,
	\label{Delta}\\
	{\cal E}_{ab}(t)&:=&\xbr\lambda_n,a;t|H|\lambda_n,b;t\xkt\;,
	\label{E}\\
	{\cal A}_{ab}(t)&:=&i\xbr\lambda_n,a;t|\frac{d}{dt}|\lambda_n,b;t\xkt\;.
	\label{A}
	\xeea
Since Eq.~(\ref{u=}) has the form of a matrix Schr\"odinger equation, its solution
can be implicitly written as
	\xbe
	u(t)={\cal T}e^{-i\int_0^t \Delta(t')dt'}u(0)=
	{\cal T}e^{i\int_0^t [-{\cal E}(t')+{\cal A}(t')]dt'}u(0)
	\;,
	\label{u=2}
	\xee
where ${\cal T}$ is the time-ordering operator. If the matrices ${\cal E}(t)=
({\cal E}_{ab})$ and ${\cal A}=({\cal A}_{ab})$ commute, then Eq.~(\ref{u=2})
can be expressed as
	\xbe
	u(t)={\cal T}e^{-i\int_0^t {\cal E}(t')dt'}
	{\cal T}e^{i\int_0^t {\cal A}(t')dt'} u(0)\;.
	\label{u=3}
	\xee

Next consider the spectral resolution of $I(t)$, namely
	\xbe
	I(t)=\sum_n \lambda_n \Lambda_n(t)\;,
	\label{I=0}
	\xee
where 
	\xbe
	\Lambda_n(t):=\sum_{a=1}^{l_n}|\lambda_n,a;t\xkt\xbr\lambda_n,a;t|=
	\sum_{a=1}^{l_n}|\lambda_n,a;t\xkt'\xbr\lambda_n,a;t|'\;
	\label{Lambda}
	\xee
is the degenerate eigenprojector (state) associated with the eigenvalue $\lambda_n$ 
or alternatively the degeneracy subspace ${\cal H}_{\lambda_n}(t)$. Now, if $I(t)$ 
is a periodic dynamical invariant with period $T$, i.e., $I(T)=I(0)$, then the 
degenerate eigenprojectors $\Lambda_n(t)$ will also be periodic, i.e., $\Lambda_n(T)
=\Lambda_n(0)$. In other words, $\Lambda_n(0)$ undergo {\em degenerate cyclic 
evolutions.}

Let us choose a set of instantaneous eigenvectors $|\lambda_n,a;t\xkt$ of $I(t)$.
Then $|\lambda_n,a;t\xkt$ are single-valued and periodic,
$|\lambda_n,a;T\xkt=|\lambda_n,a;0\xkt$. Now consider the evolution of a frame 
	\xbe
	\Psi(0)=\{ |\psi_1(0)\xkt,\cdots,|\psi_{l_n}(0)\xkt\}
	\label{frame0}
	\xee
of ${\cal H}_{\lambda_n}(0)$. We can choose the initial condition in such a way 
as $|\psi_a(0)\xkt=|\lambda_n,a;0\xkt=|\lambda_n,a;0\xkt'$. Then the vectors
constituting the frame evolve according to $|\psi_a(t)\xkt=|\lambda_n,a;t\xkt'$. 
After a complete cycle, therefore, one obtains a new frame $\Psi(T)$ of the 
degeneracy subspace  ${\cal H}_{\lambda_n}(T)={\cal H}_{\lambda_n}(0)$ which is
related to $\Psi(0)$ by the unitary transformation~(20), with $u(0)=1$, or 
alternatively by~(21)
if the matrices ${\cal E}(t)$ and ${\cal A}(t)$ commute. For $t=T$,
the second time-ordered exponential in (21) is precisely the 
non-adiabatic non-Abelian geometric phase factor of Anandan \cite{anandan}.

Note that under the single-valued unitary (gauge) transformations of
the basis vectors $|\lambda_n,a;R(t)\xkt$, the non-Abelian geometric phase
factor (20) transforms covariantly --- not invariantly. This means that the 
physically observable quatities depend only on its invariants, namely, its 
eigenvalues. As discussed in Ref.~\cite{anandan}, one can in general find 
an eigenbasis $\{\lambda_n,a;R(t)\xkt\}$ in which the total non-Abelian phase 
factor (20) is diagonal. In this basis, the invariant diagonal elements which 
are of physical importance can be written as the product of a dynamical part
and a geometrical part.

\section{Periodic Invariants and Floquet Decompositions}

As discussed in the preceding section, given a dynamical invariant $I(t)$ 
one can obtain solutions of the corresponding Schr\"odinger equation as
eigenvectors of $I(t)$, provided that one performs the necessary unitary
transformations. The converse of this procedure is also valid. In
order to see this, let ${\cal C}:=\{ |\psi_n(0)\xkt\}$ be a complete set of
state vectors, $c_n\in\xR$, and $|\psi_n(t)\xkt:=U(t) |\psi_n(0)\xkt$ 
be the solutions of the Schr\"odinger equation (\ref{sch-eq}) corresponding
to the initial conditions $|\psi_n(t=0)\xkt=|\psi_n(0)\xkt$. Then
	\xbe
	I(t):=\sum_n c_n |\psi_n(t)\xkt\xbr\psi_n(t)|
	=U(t)\left[\sum_n c_n|\psi_n(0)\xkt\xbr\psi_n(0)|\right]
	U^\dagger(t)=U(t)I(0)U^\dagger(t)\;
	\label{I-c}
	\xee
is a dynamical invariant.\footnote{Note that the numbers $c_n$ are not 
generally distinct.} In fact, every dynamical invariant $I(t)$ can be viewed
as associated with a complete set ${\cal C}$ of initial state vectors, and 
expressed as 
	\xbe
	I(t)=U(t)I(0)U^\dagger(t)\;.
	\label{uiu}
	\xee

For a periodic dynamical invariant $I(t)$ with period $T$, one has
	\xbe
	I(0)=I(T)=U(T)I(0)U^\dagger(T)=e^{iMT}I(0)e^{-iMT}\;,
	\label{i=per}
	\xee
For a generic value of $T$, this implies\footnote{There are special situations
where $\left[e^{iMT},I(0)\right]=0$ will not imply $[M,I(0)]=0$. See footnote~1
for an example. The results presented below can also be obtained using 
Eq.~(\ref{i=per}) directly.}
	\xbe
	\left[I(0),M\right]=0\;.
	\label{i,m}
	\xee
In particular $M$ and $I(0)$ have simultaneous eigenvectors, and
	\xbe
	I(t)=Z(t)I(0)Z^\dagger(t)\;.
	\label{i=ziz}
	\xee

Now let us recall that the eigenprojectors $\Lambda_n(0)$ of $I(0)$ represent
the degenerate cyclic states. If $\Lambda_n(0)$ happens to also be an eigenprojector
of $M$, then the corresponding total phase factor is Abelian. Therefore, a necessary
condition for the occurrence of a non-Abelian geometric phase is the existence of a 
$T$-periodic dynamical invariant $I(t)$ such that $I(0)$ has a degenerate 
eigenprojector which is not an eigenprojector of the Floquet operator $M$, i.e., 
$I(0)$ and $M$ have different degeneracy structures.

This observation suggests a way of obtaining degenerate cyclic states
even for systems whose Hamiltonian and (or) evolution operator are 
non-degenerate. 

\section{Application to the Spin System}

Perhaps the best-known example of a model which leads to Abelian geometric
phases is a magnetic dipole (a spin) interacting with a changing classical
magnetic field, \cite{berry1984,pra2,bohm-qm,jmp}. The Hamiltonian is given by
	\xbe
	H(t)=b \vec R(t)\cdot\vec J\;,
	\label{Hamiltonian}
	\xee
where $b$ is (proportional to) the Larmor frequency, $\vec R(t)\in R^3$ 
describes the magnetic field vector, and $\vec J$ is the angular momentum 
of the dipole. For $\vec R(t)\neq 0$, the eigenvalues of $H(t)$ are all
non-degenerate and given by $b|\vec R|k$ where $k$ is a half-integer. If
the dipole has total angular momentum $j$, then $k\in\{-j,-j+1,\cdots,j\}$.

In the following we shall restrict ourselves to the simplest possible
dipole system which would allow for a degenerate cyclic evolution.
Clearly, this is the $j=1$ case, where the Hilbert space is three-dimensional.
The general problem of non-Abelian adiabatic geometrical phase for systems 
with a three-dimensional Hilbert space is discussed in \cite{jpa}. Here we
are interested in the non-adiabatic non-Abelian phases.

In view of the results of the Floquet theory, let us consider an evolution
operator of the form (\ref{flo}) with
	\xbe
	Z(t)=e^{i\Omega t J_1}\;,~~~~M=\omega J_3\;,
	\label{z-m}
	\xee
where $\Omega:=2\pi/T$ and $\omega$ is an arbitrary positive real constant. This 
corresponds to the $T$-periodic Hamiltonian,
	\xbe
	H=-\left[\Omega J_1+\omega\sin(\Omega t)J_2+\omega\cos(\Omega t)J_3\right]
	\;,
	\label{H=}
	\xee
of a magnetic dipole interacting with a precessing magnetic field.

Clearly the pure cyclic states are eigenstates of $M=\omega J_3$. Since
$j=1$, $J_3$ has three non-degenerate eigenvalues, namely $-1,~0,$ and $1$. We
shall denote the corresponding eigenvectors by $|-\xkt,~|0\xkt$, and $|+\xkt$,
respectively. Therefore, we can write
	\xbe
	M=\omega(|+\xkt\xbr +|-|-\xkt\xbr -|)\;.
	\label{M=}
	\xee

Now consider an invariant $I(t)$ of the form $Z(t)I(0)Z^\dagger(t)$ with
	\xbe
	I(0)=\lambda_1(|+\xkt\xbr +|+|0\xkt\xbr 0|)+\lambda_2|-\xkt\xbr -|\;
	,~~~~\lambda_1,\lambda_2\in\xR-\{0\}\;.
	\label{I=}
	\xee
Obviously $I(0)$ commutes with $M$ and $I(t)$ is $T$-periodic.
Moreover, $\Lambda_1(0):=|+\xkt\xbr +|+|0\xkt\xbr 0|$ is a degenerate eigenprojector 
which undergoes a cyclic evolution.

If we choose $\Psi(0)=\{|+\xkt,|0\xkt\}$ as the initial frame for the 
cyclic evolution the unitary operation relating $\Psi(0)$ to $\Psi(T)$ will
be diagonal and the total phase factor will still be Abelian. However, if we
choose an arbitrary initial frame,
	\xbe
	\Psi(0)=\{|\psi_1(0)\xkt,\psi_2(0)\xkt\},~~~
	|\psi_1(0)\xkt:=\xi |+\xkt+\zeta|0\xkt,~~~
	|\psi_2(0)\xkt:=\zeta^* |+\xkt-\xi^*|0\xkt\;,
	\label{frame}
	\xee
with $\xi,\zeta\in\xC$ and $|\xi|^2+|\zeta|^2=1$, then we obtain a non-diagonal
unitary operator and a non-Abelian phase. Using the basic properties
of the angular momentum operators and the relation $|\lambda_1,a;t\xkt=Z(t)
|\lambda_1,a;0\xkt$, which follows from $I(t)=Z(t)I(0)Z^\dagger(t)$, we can 
easily obtain
	\xbea
	{\cal E}&=&-\omega\left(\begin{array}{cc}
	|\xi|^2  & \xi^*\zeta^*\\
	\xi\zeta & |\zeta|^2\end{array}\right)
	-\frac{\Omega}{\sqrt{2}}\left(\begin{array}{cc}
	\xi^*\zeta+\zeta^*\xi & -\xi^{*2}+\zeta^{*2}\\
	-\xi^2+\zeta^2        & -(\xi^*\zeta+\zeta^*\xi)\end{array}\right)\;,\\
	{\cal A}&=&-\frac{\Omega}{\sqrt{2}}\left(\begin{array}{cc}
	\xi^*\zeta+\zeta^*\xi & -\xi^{*2}+\zeta^{*2}\\
	-\xi^2+\zeta^2        & -(\xi^*\zeta+\zeta^*\xi)\end{array}\right)\;,\\
	\Delta&=&-\omega\left(\begin{array}{cc}
	|\xi|^2  & \xi^*\zeta^*\\
	\xi\zeta & |\zeta|^2\end{array}\right)\;.
	\xeea
Since ${\cal E},~{\cal A}$ and $\Delta$ do not depend on time, we have
	\xbe
	\Psi(T)=e^{-iT\Delta}\Psi(0)\;.
	\label{psi=}
	\xee
Note that
${\cal E}$ and ${\cal A}$ do not generally commute and 
Eq.~(21) does not hold.

\section{Conclusion}

In this article, we have explored the relationship between the Floquet
decomposition of the evolution operator for a periodic Hamiltonian and
the periodic dynamical invariants of Lewis and Riesenfeld. We showed
that the degenerate cyclic states may be viewed as the eigenstates of
a Hermitian operator $I(0)$ which serves as the initial value of a 
periodic dynamical invariant $I(t)$. We derived the expression for the
corresponding non-Abelian non-adiabatic geometrical phase in terms
of the eigenvectors of $I(t)$; thus generalizing the results of
Monteoliva, et al \cite{mkn} to degenerate cyclic evolutions.

We used our results to show that the simple quantum system describing
the dynamics of a magnetic dipole in a precessing magnetic field, which
since Berry's paper \cite{berry1984} has served as the main example 
involving a nontrivial Abelian geometrical phase, also leads to degenerate
cyclic evolutions.

We wish to conclude this paper with the following remarks:
	\begin{itemize}
	\item[1)] The Moore-Stedman \cite{m-s} method of generating the
pure cyclic states as eigenstates of the Floquet operator $M$ may be 
viewed as a special case of the method of invariants. 
This can be easily seen, by associating a $T$-periodic dynamical invariant,
namely $I(t):=Z(t)MZ^\dagger(t)$, to a given Floquet decomposition 
$Z(t)e^{iMt}$ of the evolution operator. 
	\item[2)] The method of invariants is superior to the method of 
Moore and Stedman \cite{m-s}, for it can also be used to generate
degenerate cyclic states.
	\item[3)] For a general quantum system with a three-dimensional
Hilbert space, one can easily apply the method developed in Ref.~\cite{jpa}
to obtain the general form of a Hermitian invariant which leads to a degenerate
cyclic evolution. In particular, one can show that the parameter space of
such invariants has the manifold structure of $\xC P^2$.
	\end{itemize}


\newpage

\begin{thebibliography}{99}
\bibitem{berry1984} M.~V.~Berry, Proc.\ Roy.\ Soc.\ London 
{\bf A 392}, 45 (1984).
\bibitem{simon} B.~Simon, Phys.~Rev.~Lett.\ {\bf 51}, 2167, (1983).
\bibitem{wi-ze} F.~Wilczek and A.~Zee, Phys.\ Rev.\ Lett.\ {\bf 52},
2111 (1984).
\bibitem{hannay} J.~H.~Hannay, J.~Phys.~A: Math.\ Gen.\ {\bf 18}, 
221 (1985);\\
M.~V.~Berry, J.~Phys.~A: Math.\ Gen.\ {\bf 18}, 15-27 (1985);\\
J.~Anandan, Phys.~Lett.~A {\bf 129}, 201 (1988).
\bibitem{aa} Y.~Aharonov and J.~Anandan, Phys.~Rev.~Lett.\ {\bf 58},
1593 (1987);\\
J.~Anandan and Y.~Aharonov, Phys.~Rev.~{\bf D38}, 1863 (1988).
\bibitem{noncyclic}  J.~Samuel and R.~Bhandari, Phys.~Rev.~Lett.\ {\bf 60},
2339 (1988);\\
J.~Zak, Europhys.~Lett., {\bf 9}, 615 (1989);\\
G.~G.~de Polavieja and E.~Sj\"oqvist, Am.\ J.\ Phys.\ {\bf 66},
431 (1998);\\
G.~G.~de Polavieja, Phys.~Rev.~Lett.\ {\bf 81}, 1 (1998);\\
A.~K.~Pati, ``Adiabatic Berry Phase and Hannay Angle for Open Paths,''
quant-ph/9804057.
\bibitem{applications} A.~Shapere and F Wilczek, eds., {\em Geometric 
Phases in Physics} (World Scientific, Singapore, 1989);\\
J.~W.~Zwanziger, M.~Koenig, and A.~Pines, Annu.\ Rev.\ Phys.\ Chem.\
{\bf 41}, 601 (1990);\\
D.~J.~Moore, Phys.~Rep., {\bf 210}, 1 (1991);\\
C.~A.~Mead, Rev.\ Mod.\ Phys.\ {\bf 64}, 51 (1992);\\
R.~Bhandari, Phys.~Rep., {\bf 281}, 1 (1997).
\bibitem{math} D.~Page, Phys.~Rev.~A {\bf 36}, 3479 (1987);\\
J.~Anandan and L.~Stodolsky, Phys.~Rev.~D {\bf 35}, 2597 (1987);\\
J.~E.~A.~Avron, L.~Sadun, J.~Segert, and B.~Simon, Commun.\ Math.\ Phys.\
{\bf 124}, 124 (1989);\\
J.~Anandan, Phys.~Lett.~A {\bf 147}, 3 (1990);\\
Sh.-J.~Wang, Phys.~Rev.~A {\bf 42}, 5103 (1990);\\
A.~Bohm, L.~J.~Boya, A.~Mostafazadeh, and G.~Rudolph, J.~Geom.~Phys.\
{\bf 12}, 13 (1993);\\
A.~Bohm and A.~Mostafazadeh, J.~Math.~Phys.\ {\bf 35}, 1463 (1994);\\
A.~Mostafazadeh, J.~Math.~Phys.\ {\bf 37}, 1218 (1996);\\
M.~A.Aguilar and M.~Socolovsky, Int.\ J.\ Theo.\ Phys.\ {\bf 36}, 
883 (1997).
\bibitem{older} M.~Born and V.~Fock, Zeit.\ F.~Phys.\ {\bf 51},
165 (1928);\\
J.~H.~Van Vleck, Phys.\ Rev.~{\bf 33}, 467 (1929);\\
T.~Kato, J.~Phys.~Soc.~Jpn.\ {\bf 5}, 435 (1950).
\bibitem{le-ri} H.~R.~Lewis Jr. and W.~B.~Riesenfeld, J.~Math.~Phys.\
{\bf 10}, 1458 (1969).
\bibitem{morales} D.~A.~Morales, J.~Phys.~A: Math.\ Gen.\ {\bf 21}, 
L889 (1988).
\bibitem{mkn} D.~B.~Monteoliva, H.~J.~Korsch and J.~A.~N\'u$\tilde{\rm n}$es,
J.~Phys.~A: Math.\ Gen.\ {\bf 27}, 6897 (1994).
\bibitem{m-s} D.~J.~Moore and G.~E.~Stedman, J.~Phys.~A: Math.\ Gen.\ {\bf 23}, 
2049 (1990).
\bibitem{moore} D.~J.~Moore, J.~Phys.~A: Math.\ Gen.\ {\bf 23}, L665 (1990).
\bibitem{fur} G. B. Furman, J.~Phys.~A: Math. \ Gen.\ {\bf 27}, 6893 (1994).
\bibitem{anandan} J.~Anandan, Phys.~Lett.~A {\bf 133}, 171 (1988).
\bibitem{floquet} G.~Floquet, Ann.~E.~N.~S.\ {\bf 12}, 47 (1883).
\bibitem{pra2} A.~Mostafazadeh,  Phys.~Rev.~{\bf A 55}, 1653 (1997).
\bibitem{bohm-qm} A.~Bohm, {\em Quantum Mechanics:
Foundations and Applications,} third edition (Springer-Verlag, Berlin,
1993).
\bibitem{jmp} A.~Mostafazadeh, J.~Math.~Phys.\ {\bf 38}, 3489 (1997).
\bibitem{jpa} A.~Mostafazadeh, J.~Phys.~A: Math.\ Gen.\ {\bf 30}, 7525 
(1997).
\end{thebibliography}
\end{document}